\documentstyle[12pt,a4wide,epsf]{article}
\begin{document}

\begin{flushright}
\today \\
hep-th/0411208 \\
\end{flushright}
\vskip 1cm
\begin{center}
{\large \bf Cosmic Acceleration in Massive Half--Maximal Supergravity}
\end{center}
\vspace*{5mm} \noindent

\centerline{Ph.~Brax\footnote{brax@spht.saclay.cea.fr}${}^{a}$,
C. van de Bruck\footnote{C.vandeBruck@sheffield.ac.uk}${}^{b}$ and 
A. C. Davis\footnote{a.c.davis@damtp.cam .ac.uk}${}^{c}$}
\vskip 0.5cm \centerline{${a}$) \em Service de Physique
Th\'eorique} \centerline{\em CEA/DSM/SPhT, Unit\'e de recherche
associ\'ee au CNRS,} \centerline{\em CEA-Saclay F-91191 Gif/Yvette
cedex, France.} \vskip 0.5cm
\vskip 0.5cm \centerline{${b}$) \em Department of Applied Mathematics}
\centerline{\em The University of Sheffield}
\centerline{\em Hounsfield Road, Sheffield S3 7RH, United Kingdom} \vskip 0.5cm

\vskip 0.5cm \centerline{${c}$) \em Department of Applied Mathematics and Theoretical Physics}
\centerline{\em Centre for Mathematical Sciences}
\centerline{\em Cambridge CB2 0WA, United Kingdom} \vskip 0.5cm

\begin{center}
{\bf Abstract}
\end{center}
We consider massive half--maximal supergravity in $(d+3)$ dimensions and
compactify it on a symmetric three--space. We find that the static 
configurations of
 $\hbox{Minkowski}_{d}\times S^3$  obtained by balancing the positive scalar 
potential for the dilaton and the flux of a three-form through the 
three--sphere are unstable. The resulting cosmological evolution 
breaks supersymmetry and leads to an accelerated expansion in $d$ dimensions.

\newpage

\section{Introduction}

Recent cosmological observations seem to indicate that the
expansion of the Universe is accelerating. The acceleration of the
expansion has been corroborated using the results of the WMAP
satellite on the anisotropies of the CMB, the Hubble diagram of
type Ia supernovae and the large scale structures of the Universe.
This is strong evidence in favour of the existence of a dark
energy fluid which may  be the realization of a cosmological
constant (see e.g. \cite{dereview} for a recent review).
The required energy density of the dark
energy fluid is some 120 orders of magnitude below a natural scale
such as the Planck mass. A satisfactory explanation for the
existence and the smallness of the cosmological constant has not
been found yet.

An accelerated phase  in the history of the Universe is also
advocated to have existed in the early Universe during inflation 
\cite{liddle}. Hence the Universe would have undergone at least
two phases of accelerated expansion. From the point of view of
high energy physics and in particular string theory, the existence
of accelerated universes is problematic \cite{town,bala}. First of
all, there is no known formulation of string theory in a
space--time with a future cosmological event horizon like de
Sitter space (see e.g. \cite{witten} and \cite{kaloper}). In a
nutshell, this springs from the difficulty of formulating S-matrix
amplitudes in de Sitter space. Of course, since string theory is
valid at very high energy well before the energy scales when the
recent acceleration of the expansion of the Universe started, this
might not be relevant to the cosmological constant problem. On the
contrary, one may hope to describe the acceleration of the
expansion within the realm of effective field theories as deduced
by compactification of  the low energy supergravity theories
associated with string theories. At the level of a four dimensional
description, natural candidates which may trigger the accelerated
expansion can be readily identified with the various moduli
arising from the compactification process \cite{quev1,quev2,gar,towo,ohta, wohl,kal1,kal2}.

Acceleration can arise when a four dimensional theory violates the
strong energy condition stating that $p\ge -\rho/3$ where $p$ is
the pressure and $\rho$ the energy density. This can be realised
with a slow rolling scalar field as well-known in inflation
models. However it turns out that string theory and M--theory in
10 and 11 dimensions do not violate the strong energy condition.
Upon compactification on a static manifold, the resulting 4d model
satisfies the strong energy condition. Therefore no static
compactification of string theory or M--theory can lead to an
accelerated Universe. This is the Gibbons--Maldacena--Nunez
theorem \cite{gib, nun}. Of course, one may circumvent the
stringent constraint of having a static compactification by
allowing a space--time dependence of the breathing mode measuring
the size of the internal manifold. In that case, one can deduce
that the effective potential for the breathing mode cannot have a
stationary point with a positive value of the potential. In
particular, runaway potentials of the exponential
form \cite{neu1,neu2}
\begin{equation}
V=\Lambda e^{-2 c \phi}
\end{equation}
are allowed. It is well--known that this leads to power law
inflation provided \cite{pow,hal}
\begin{equation}
c<1/\sqrt{2}.
\end{equation}
It turns out the  compactifications with fluxes or with an
hyperbolic geometry do not lead to $c<1/\sqrt{2}$. In the same
vein, the massive type IIA supergravity compactified on a six
torus leads to $c=\sqrt 7$ \cite{town}.  Recently it was
conjectured that compactifications leading to $c<1/\sqrt 2$ are
not allowed in string theory \cite{town}. In the following we will
consider a compactification of heterotic string theory on a torus
times a circle \cite{pope}. The torus plays no role here, the
compactification along the circle uses a gauged symmetry of the
equations of motion. A consistent truncation of such a
$(d+4)$--dimensional theory results in a massive supergravity
theory with a positive potential \cite{pope}. Moreover, a scalar
field combining the radius of the circle and the dilaton received
a positive exponential potential with an exponent
$c=\frac{1}{\sqrt {10}}<\frac{1}{\sqrt{2}}$. The resulting theory
possesses static solutions in the form of Minkowski space times a
3--sphere traversed by the  non--zero flux of a three--form. We
show that these static and supersymmetric configurations are
unstable. We study the cosmological dynamics of the model and find
that cosmological solutions in the form of an accelerating $d$
space--time times a sphere whose size grows with time can be
found. This gives an example of a compactification of string
theory/supergravity  leading to an accelerating Universe.

\section{Massive Half--Maximal Supergravity}

Here we follow closely the paper \cite{pope} where more details
can be found. Let us start with $(d+4)$ dimensional half--maximal
supergravity, i.e. with sixteen supersymmetries as in heterotic string theory. The bosonic
field content is as follows. There is
gravity $\hat g_{\mu\nu}$, the antisymmetric tensor $\hat
B_{\mu\nu}$, the dilaton $\hat \phi$ and  $(6-d)$
vector fields $\hat A^a_\mu$. The $(d+4)$ dimensional action reads
\begin{equation}
{\cal S}_{d+3}= \int (\hat R*1-\frac{1}{2} *d\hat \phi \wedge
d\hat \phi -\frac{1}{2} e^{\alpha \hat \phi} *\hat H\wedge \hat H
-\frac{1}{2} e^{\frac{1}{2} \alpha  \hat \phi}*\hat F^a\wedge \hat
F^a).
\end{equation}
where $* 1$ is the volume form and $\alpha^2= \frac{8}{d+2}$. We
have defined $\hat F^a=d\hat A^a$ and $\hat H= d\hat B -
\frac{1}{2} \hat F^a \wedge \hat A^a$. The equations of motion are
invariant under the two transformations
\begin{equation}
\hat \phi \to \hat \phi +\frac{1}{\alpha} \lambda _1, \ \ d\hat
s^2\to e^{2\lambda_2} d\hat s^2
\end{equation}
and
\begin{equation}
\hat B\to e^{-2\lambda_1 + \lambda _2} \hat B, \ \ \hat A^a \to
e^{-\lambda_1 +\lambda_2} \hat A^a.
\end{equation}
The next step consists in dimensionally reducing  to $(d+3)$
dimensions around a circle $S^1$. This is achieved via the
decomposition
\begin{eqnarray}
d\hat s^2&= &e^{mz}(e^{2\beta \psi} ds^2 + e^{2\gamma \psi}(dz +
\tilde A)^2),\quad \hat B= B+ \tilde B \wedge dz\nonumber \\
\hat A^a&=& A^a + \xi^a dz,\quad \hat \phi = \tilde \phi+
\frac{4}{\alpha} mz,\nonumber \\
\end{eqnarray}
where $z$ is the coordinate around the circle $S^1$ and
\begin{equation}
\beta^2=\frac{1}{2(d+1)(d+2)},\quad \gamma=-(d+1)\beta.
\end{equation}
The theory reduces to a half--maximal supergravity theory coupled
to a vector multiplet and  can be further truncated by
putting
\begin{equation}
\tilde B=0, \quad \tilde A=0,\quad \xi^a=0, \quad A^a=0,
\end{equation}
and choosing
\begin{equation}
\psi=-\frac{4\beta}{\alpha} \tilde\phi.
\end{equation}
Redefining now
\begin{equation}
\phi= \frac{2\alpha}{\tilde a} \tilde \phi,
\end{equation}
where $\tilde a=\sqrt{\frac{8}{d+1}}$, the $(d+3)$ --bosonic
dynamics reduce to a field theory whose Lagrangian reads
\begin{equation}
{\cal L}= R -\frac{1}{2} (\partial \phi)^2 -\frac{1}{4} e^{\tilde
a \phi} H^2 -(d-1)^2 m^2 e^{-\frac{\tilde a}{2} \phi}.
\end{equation}
First of all, the end result is a massive supergravity theory
with a positive potential. Moreover,  notice that for $d=4$, the
positive potential has an exponent
\begin{equation}
c\equiv \frac{\tilde a}{4} =  \sqrt{\frac{1}{10}}, \quad d=4,
\end{equation}
contradicting the conjecture presented in \cite{townsend}. We will now analyse the
dynamics of this theory. It presents an interesting interplay
between the potential term and the 3--form term leading  to
unstable static configurations. These cosmological solutions break
supersymmetry.

\section{Cosmological spacetimes}

From the action, the equations of motion can be found to be
\begin{eqnarray}
R_{\mu\nu} &=& \frac{1}{2}\partial_\mu \phi \partial_\nu \phi +
\frac{1}{4}e^{\tilde{\alpha}\phi}\left(H_{\mu\rho\sigma}H_{\nu}^{~\rho\sigma}
- \frac{2}{3(d+1)}H^2_{(3)} g_{\mu\nu} \right) \nonumber \\
&+& \frac{m^2 (d+2)^2}{d+1}e^{-\frac{1}{2}\alpha\phi}g_{\mu\nu}, \\
\Box \phi &=& \frac{e^{\alpha\phi}}{3\sqrt{2(d+1)}}H^2_{(3)}
- \frac{\sqrt{2}(d+2)^2 m^2}{\sqrt{d+1}} e^{-\frac{1}{2}\alpha\phi},\\
\nabla^\rho \left(e^{\tilde{a}\phi} H_{\mu\nu\rho} \right) &=& 0.
\end{eqnarray}
We will specialise these equations and only consider time dependent solutions, i.e. we are only interested
in cosmological solutions.
For the metric we choose the ansatz
\begin{equation}
ds^2 = a^2 (t)\eta_{ab} dx^adx^b + b^2(t) g_{ij} dx^i dx^j ,
\end{equation}
where Roman letters at the beginning of the alphabet denote the
cosmological $d$ space--time and  $g_{ij}$ is a metric of constant
curvature $k$. Here the two scale factors $a$ and $b$ are
independent. Notice that the size of the 3--sphere is allowed to
vary in time. For the field $H$ we use
\begin{equation}
H_{ijk}(t) = f(t)\epsilon_{ijk},
\end{equation}
and zero otherwise.
It corresponds to a net flux across the 3--space of curvature $k$. As long as $f$ is not constant, the
flux  varies in time.  The equations of motion can be found to be
\begin{eqnarray}
-\left(d - 1 \right)\dot {\cal H} &=& \frac{1}{2}\dot \phi^2
+ \frac{C}{d+1}a^2
- \frac{a^2 m^2 (d+2)^2}{d+1} e^{-\frac{1}{2}\tilde{a}\phi} \label{Friedmanndot}\\
\frac{{\cal H}^2}{a^2} &=& \frac{1}{2(d-1)(d-2)}\left(\frac{\dot \phi}{a}\right)^2
- \frac{C}{d^2 - 1}
+ \frac{(d+2)^2}{d^2 - 1} m^2 e^{-\frac{1}{2}\tilde{a}\phi} \label{Friedmann}\\
\ddot \phi + (d-2){\cal H}\dot \phi + 3 \frac{\dot b}{b}\dot \phi &=&
\frac{\sqrt{2}(d+2)^2 m^2}{\sqrt{d+1}} a^2 e^{-\frac{1}{2}\tilde{a}\phi}
- \sqrt{\frac{2}{d+1}}C a^2 \\
\frac{\ddot b}{b} + 2 \left( \frac{\dot b}{b} \right)^2 &+& (d-2) {\cal H} \frac{\dot b}{b}
 + \frac{2k a^2}{b^2} = \frac{a^2}{2}\frac{d-1}{d+1} C
+ \frac{m^2 (d+2)^2}{d+1} a^2 e^{-\frac{1}{2}\tilde{a}\phi}
\end{eqnarray}
in the cosmological context.
In these equations, we have defined ${\cal H} = a'/a$ (prime denotes derivative with respect to
conformal time). The quantity $C$ is defined by
\begin{equation}\label{fluxansatz}
f^2(t) = C(t) b(t)^6 e^{-\tilde{a}\phi}.
\end{equation}
Note that this implies, that $C$ is always positive. This ansatz couples the scalar field $\phi$,
the scale factor $b$ and $H_{ijk}$ in a particular manner. Consistency between the field equations
requires that $C$ fullfills
\begin{eqnarray}
\frac{d-2}{d^2 - 1} a^2 \dot C &=& -\frac{3}{d-1}\left(\frac{\dot b}{b}\right)\dot\phi^2
+ \frac{\sqrt{2}(d+2)^2 m^2}{(d-1)\sqrt{d+1}} a^2 \dot \phi e^{-\frac{1}{2}\tilde{a}\phi} \nonumber \\
&-& \frac{\sqrt{2} a^2}{(d-1)\sqrt{d+1}} C \dot \phi
- \frac{\tilde{a}}{2}\dot\phi \left(\frac{(d-2)(d+2)^2}{d^2 - 1}\right)
a^2 m^2 e^{-\frac{1}{2}\tilde{a}\phi},
\end{eqnarray}
which provides an equation for the evolution of the three-form field.
In cosmic time, the equations read
\begin{eqnarray}
H^2 &=& \frac{1}{2(d-1)(d-2)}\dot \phi^2
- \frac{C}{d^2 - 1}
+ \frac{(d+2)^2}{d^2 - 1} m^2 e^{-\frac{1}{2}\tilde{a}\phi} \label{Friedmann1}\\
\ddot \phi + (d-1)H\dot \phi + 3 \frac{\dot b}{b}\dot \phi &=&
\frac{\sqrt{2}(d+2)^2 m^2}{\sqrt{d+1}} e^{-\frac{1}{2}\tilde{a}\phi}
- \sqrt{\frac{2}{d+1}}C  \\
\frac{\ddot b}{b} + 2 \left( \frac{\dot b}{b} \right)^2 &+& (d-1) H \frac{\dot b}{b}
 + \frac{2k}{b^2} = \frac{1}{2}\frac{d-1}{d+1} C
+ \frac{m^2 (d+2)^2}{d+1} e^{-\frac{1}{2}\tilde{a}\phi} \\
\frac{d-2}{d^2 - 1} \dot C &=& -\frac{3}{d-1}\left(\frac{\dot b}{b}\right)\dot\phi^2
+ \frac{\sqrt{2}(d+2)^2 m^2}{(d-1)\sqrt{d+1}} \dot \phi e^{-\frac{1}{2}\tilde{a}\phi} \nonumber \\
&-& \frac{\sqrt{2}}{(d-1)\sqrt{d+1}} C \dot \phi
- \frac{\tilde{a}}{2}\dot\phi \left(\frac{(d-2)(d+2)^2}{d^2 - 1}\right)
m^2 e^{-\frac{1}{2}\tilde{a}\phi}.
\end{eqnarray}
In these equations, a dot denotes a derivative with respect to cosmic time and $H = \dot a/a$.

Let us first analyse static solutions. It is easy to see that static backgrounds are
given by
\begin{equation}
C_0= (d+2)^2 m^2 e^{-\frac{\tilde a}{2}\phi_0}.
\end{equation}
Such static configurations are only possible for
\begin{equation}
k=1,
\end{equation}
corresponding to a spherical compactification with
\begin{equation}
b_0= \frac{2}{\sqrt C_0}.
\end{equation}
Moreover the resulting configurations are  known to be supersymmetric. Indeed, they are the dimensional reductions of
the near horizon limits of $(d+1)$ branes \cite{pope}.

Let us investigate the stability and consider small (i.e. linear) fluctuations around the static solution above.
The fluctuation $\delta\phi$ behaves
like a massless field, i.e.
\begin{equation}
(\delta \phi)^{..} = 0
\end{equation}
Hence the field $\phi$ possesses a flat direction around the static configuration. Similarly
the scale factor $a$ fullfills the same type of equation,
\begin{equation}
(\delta a)^{..} = 0
\end{equation}
and the $d$--dimensional space--times remains static.
 On the other hand, fluctuations in $C$ are determined by fluctuations in $\phi$:
\begin{equation}
\delta C = - \sqrt{\frac{2}{d+1}} m^2 (d+2)^2 e^{-\frac{1}{2}\tilde{a}\phi_0} \delta \phi.
\end{equation}
From these equations, one can find that fluctuations in $b$ around the static background are governed by
\begin{equation}
(\delta b)^{..} = \frac{m^2 (d+2)^2}{2}e^{-\frac{1}{2}\tilde{a}\phi_0}\left[ \delta b
- \sqrt{\frac{2}{5}}\delta\phi \right],
\end{equation}
which signals an exponential instability for $b$, sourced by the field fluctuation $\delta\phi$.
These considerations tell us, that fluctuations quickly become non--linear and therefore the
subspace defined by the metric $\tilde{g}_{ij}=b(t)g_{ij}$ is unstable.
To go beyond the linear perturbation analysis, we resort to a numerical study
in which we focus on $d=4$.

\begin{figure}
\epsfxsize=6cm
\epsfysize=6cm
\begin{center}
\leavevmode
\epsffile{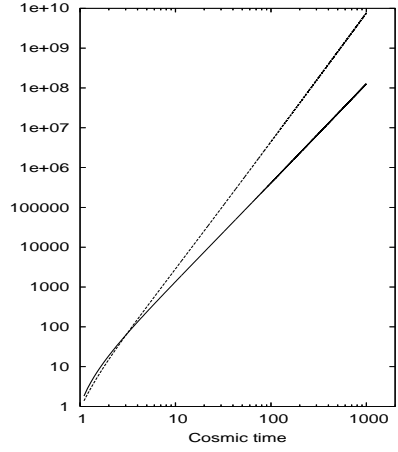}
\end{center}
\caption{Evolution of the scale factors $a(t)$ (solid line) and $b(t)$ (dashed line) as a function
of cosmic time $t$. In this example we have chosen $m=6$. The initial conditions for $\phi$ and $C$
are given by $\phi_{in} = 1.0$ and $C_{in} = 10$ (in natural units).}
\end{figure}

\begin{figure}
\epsfxsize=6cm
\epsfysize=6cm
\begin{center}
\leavevmode
\epsffile{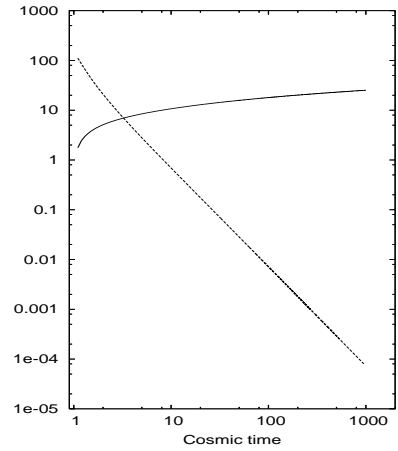}
\end{center}
\caption{Evolution of $\phi(t)$ (solid line) and $C(t)$ (dashed line) as a function
of cosmic time $t$. The parameters and initial conditions are chosen as in Fig. 1.}
\end{figure}

A typical example is given in Figure 1 and Figure 2. The scale factor
$b$ grows faster than the scale factor $a$ and $C$ decays. This holds also for other
initial conditions and parameters than those used in Figure 1 and Figure 2. The effects of
$m$, $C_0$ and $\phi_0$ are such  they only  affect  the initial behaviour of $a$, $b$,
$C$ and $\phi$. However, after a certain amount of time the behaviour of the fields is similar
to the one showed in the Figures. In particular, it is noticeable that the size of the sphere
grows with time
and is not bounded from above. In the future the size of the sphere becomes of the order of the
Hubble radius of the
4d space--time implying that, in a sense, space--time  decompactifies.

This is not the only peculiar feature of the model. Let us now turn to the physics as seen by
test matter.
Let us consider that matter is only present in 4d, and therefore corresponds to an action
\begin{equation}
S_m= \int d^4 x {\cal L}_m (\psi_m, g_{ab}),
\end{equation}
where $\psi_m$ is a 4d matter field coupled to the metric $g_{ab}=a^2\eta_{ab}$.
Let us now consider the Einstein--Hilbert term after dimensional reduction and integration
over the 3--sphere
\begin{equation}
\int d^7x R\supset \int d^4x b^3 R_{(4)},
\end{equation}
where the volume of the 3-sphere has been normalised to unity and $R_{(4)}$ is the curvature
of the metric $g_{ab}$.
Now the gravitational constant is time dependent, involving a $b^3$ factor.
One can go to the Einstein frame by defining
\begin{equation}
g_{ab}^E= b^3 g_{ab}.
\end{equation}
In this frame, Newton's constant is time independent while the coupling to matter reads
 \begin{equation}
S_m= \int d^4 x {\cal L}_m (\psi_m, b^{-3}g^E_{ab}).
\end{equation}
In particular, the effective scale factor is
\begin{equation}
a_E= b^{3/2}a.
\end{equation}
which grows faster than $t$, i.e. leads to an accelerated expansion (see Figure 3).

\begin{figure}
\epsfxsize=6cm
\epsfysize=6cm
\begin{center}
\leavevmode
\epsffile{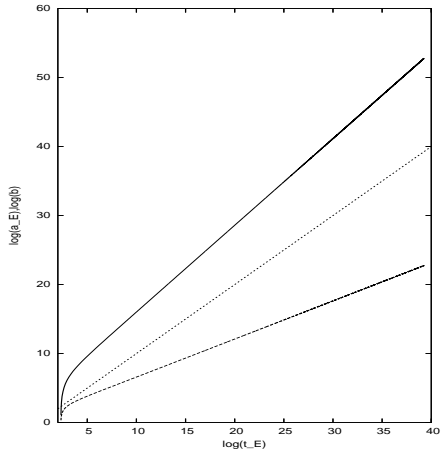}
\end{center}
\caption{Evolution of the Einstein frame scale factor $a_E$ (solid line) and $b$ (long dashed line)
as a function of time in the Einstein frame $t_E$. The parameters and
initial conditions are chosen as in Fig. 1. The short dashed line shows the line in which
$a_E \propto t_E$. Thus, in the effective four-dimensional
theory the scale factor grows faster than $t_E$, i.e. the universe is accelerating. The scale factor
$b$ grows slower than $t_E$.}
\end{figure}

In the low--energy effective theory, the coupling to to matter is not minimal anymore
but involves the factor $b^{-3}$.
In the Einstein frame, particles move no longer on geodesics and their masses are no
longer constant: massive particles of mass $m_G$ are subject to a force
$F_{\mu}= -m_G \partial_\mu \ln b^{-3/2}$ whose only non-vanishing component is
\begin{equation}
F_0=\frac{3}{2}m_G{\cal H}_b,
\end{equation}
where ${\cal H}_b= \frac{\dot b}{b}$.
The time variation of masses is given by
\begin{equation}
\frac{\dot m_G}{m_G}= -3 {\cal H}_b.
\end{equation}
This seems to result in a very large variation which might prevent the use of the model for
late time acceleration. To really answer this question one needs to take into account
the back--reaction of matter on the expansion of the Universe. This, however, 
is beyond the scope of this 
paper and left for future work.

\section{Conclusions}
We have seen in this paper that the supergravity theory presented
in \cite{pope} can lead to an accelerated expansion of the
universe. Although the theory presented is not consistent with
standard cosmology, as it predicts a huge variation of masses
during cosmic history, it might be a starting point for future
investigations. For example, we have not studied brane sources in
the theory as well as more complicated compactification schemes.
In the present case, the cosmological solutions and the
instability of the supersymmetric configurations result from the
required fine--tuning to obtain a static and supersymmetric
solution when compactifying on a three--sphere. The flux through
the compact three--sphere can only compensate the positive and
run--away potential for the dilaton when the flux is fine--tuned.
Away from this fine--tuned value, the system is unstable and the
dilaton starts running under the influence of the breathing mode
measuring the size of the three--sphere. The fact that the
instability is triggered by the breathing mode is in accord with a
violation of one of the premises of the Gibbons-Maldacena--Nunez
theorem, i.e. the requirement of a static compactification. It
remains to be seen whether such a model may have some
phenomenological applications in the presence of matter.

\vspace{1cm}

{\bf Acknowledgements:} We are grateful to C. Pope and F. Quevedo for useful 
comments. This work is supported in part by a British Council--Alliance
exchange grant, by PPARC (CvdB and ACD) and the E.U.R.T.N. "The quest for unification:
theory confronts experiment" MRTN-CT-2004-503369 (Ph. B.).

\end{document}